\documentclass[10pt, preprint]{emulateapj}		%change back to {aastex} upon submission
\usepackage{color}
\usepackage{style_wei_symbols}
\usepackage{style_wei}

\usepackage{natbib}
\begin{document}

\title{First {\it SDO} AIA Observations of a Global Coronal EUV ``Wave":
Multiple Components and ``Ripples"}

	%First Joint {\it SDO} AIA and {\it STEREO} Observations
	% Multiple Ripples and Fast Components}
	%New Insights to Global Coronal EUV Waves: First Observations by the {\it Solar Dynamics Observatory} ({\it SDO})
	%Atmospheric Imager Array (AIA)}

 	%First Double Quadrature Observations by {\it SDO} AIA and {\it STEREO} EUVI
	%Atmospheric Imager Array 

\author{Wei Liu\altaffilmark{1}\altaffilmark{2}, Nariaki V.~Nitta\altaffilmark{1}, Carolus J.~Schrijver\altaffilmark{1}, Alan M.~Title\altaffilmark{1},
 and Theodore D.~Tarbell\altaffilmark{1}}

	%Richard A.~Shine\altaffilmark{1}

\altaffiltext{1}{Lockheed Martin Solar and Astrophysics Laboratory, Department ADBS, 
Building 252, 3251 Hanover Street, Palo Alto, CA 94304}
\altaffiltext{2}{W.~W.~Hansen Experimental Physics Laboratory, Stanford University, Stanford, CA 94305}

\shorttitle{First Global EUV ``Wave" Observed by {\it SDO} AIA}
\shortauthors{Liu et al.}
\slugcomment{The Astrophysical Journal Letters, 723: L53­-L59, 2010 November 1}
	%\slugcomment{Received 2010 July 30; accepted 2010 September 20; published 2010 October 11}
	%\slugcomment{Accepted by ApJ Letterts, September 20, 2010}

\begin{abstract}	% 250 (300) words limit by ApJL (ApJ)

We present the first \sdo AIA observations of a global coronal EUV disturbance (so-called ``EIT wave")
	%involving multiple temperatures and spatio-temporal scales
revealed in unprecedented detail.
	%--- group points from summary ---
	%--- wave components ---
The disturbance observed on 2010 April 8 exhibits two components: one {\it diffuse pulse} 	%of emission enhancement
	%superimposed by %in addition 
superimposed on which are multiple {\it sharp fronts} that have slow and fast components.
	%--- CME driven ---
The disturbance originates in front of erupting coronal loops and some sharp fronts undergo accelerations, 
both effects implying that the disturbance is driven by a CME.
	%--- real wave nature: ripples, overtaking ---
The diffuse pulse, propagating at a uniform velocity of 204--$238 \kmps$
with very little angular dependence within its extent in the south, 
maintains its coherence and stable profile for $\sim$30~minutes.
Its arrival at increasing distances coincides		% seems to trigger 
with the onsets of loop expansions and the slow sharp front.
The fast sharp front overtakes the slow front, producing multiple ``ripples" and steepening the local pulse,
and both fronts propagate independently afterwards. This behavior resembles the nature of real waves.
	%--- wave profiles ---
Unexpectedly, the amplitude and FWHM of the diffuse pulse decrease linearly with distance. 
	%--- heating ---
	%The diffuse pulse appears as emission enhancement at hotter 193~\AA\ but reduction at cooler 171~\AA,
	%suggestive of heating, while the sharp fronts appear as enhancement at both wavelengths, indicating density increase.
	%--- dimming ---
	%--- model/interpretation ---
A hybrid model,	% (e.g. Cohen et al.), 
combining both wave and non-wave components, 	%is the most promising and
can explain many, but not all, of the observations.
	%--- fast feature ---
	%In addition, we find fast (600--$1100 \kmps$) features repeated at $\sim$100~s intervals as tentative evidence of {\bf fast mode} MHD waves.
Discoveries of the 	%fast features
two-component fronts and multiple ripples were made possible for the first time thanks to AIA's high cadences ($\le $20~s)
and high signal-to-noise ratio.

\end{abstract}

	%ApJ requires <6 keywords	Sun: coronal mass ejections: CMEs

%======================================================================================
\section{Introduction}
\label{sect_intro}

	%--- this paragraph can be cut to save space, go straight to EUV waves ---------
	%--- global wave obs: history review, multiple wavelength (Vrsnak & Cliver 2008---
%{\it Global} waves or traveling disturbances 
Expanding annular, global-scale, coronal EUV enhancements were discovered with \soho EIT
	%the EUV Imaging Telescope \citep[EIT;][]{SOHO.EIT.1995SoPh..162..291D} on board \soho 
and have been often called ``EIT waves"
\citep{Moses.EIT-wave.1997SoPh..175..571M, ThompsonB.EIT-wave-discover.1998GeoRL..25.2465T}.
They propagate across the solar disk at typical velocities of 200--$400 \kmps$ 
\citep{ThompsonB.EIT-wave-catalog.2009ApJS..183..225T},		%[range: 50--$700 \kmps$;][]
slower than \citet{Moreton.wave.1960AJ.....65U.494M} waves ($\sim$1000 $\kmps$).
%
	%--- focus on EUV wave: 1) obs summary - models: 1) origin 2) nature ---
Their nature is still under intense debate.
Competing models include: 	%for the nature of EUV waves include:
 {\it fast-mode} MHD waves \citep{	%ThompsonB.EIT-Moreton-wave.1999ApJ...517L.151T,
Wills-DaveyThompson.TRACE-EUV-wave.1999SoPh..190..467W,
WangYM.EIT-fastMHDwave.2000ApJ...543L..89W,	
Wu.EIT-fast-MHD-wave.2001JGR...10625089W, WarmuthA.EIT-Moreton-wave-fast-mode.2001ApJ...560L.105W, 
OfmanThompson.EIT-wave-fast-mode.2002ApJ...574..440O, Schmidt.Ofman.3D-MHD-eitwv.2010ApJ...713.1008S},
 {\it slow-mode} waves or solitons \citep{Wills-Davey.EIT-wave-soliton.2007ApJ...664..556W, 
WangHongjuan.EIT-slow-mode-wave.2009ApJ...700.1716W},
and {\it non-waves} 	%pseudo waves or propagating disturbances 
related to a current shell or successive 	%stretching or 
restructuring of field lines at the CME front 
\citep{Delannee.EIT-pseudo.wave.2000ApJ...545..512D, 
	%Delannee.EIT-wave-current-shell.2008SoPh..247..123D,
ChenFP.EIT-wave-MHD.2002ApJ...572L..99C, ChenFP.EIT-wave.2005ApJ...622.1202C, 
Attrill.EIT-wave-CME-Footprint.2007ApJ...656L.101A, Attrill.dispute-Gopal-reflection.2010ApJ...718..494A}.		%, DaiY.EUVI-wave.non-wave.2010ApJ...708..913D}.
	%or to a current shell between an expanding CME flux rope and the surrounding
	%field \citep{Delannee.EIT-wave-current-shell.2008SoPh..247..123D}.
	%\citet{CohenO.EITwave.non-wave.both.2009ApJ...705..587C} suggested that 
	%``both wave and non-wave models are required to explain the observations".
Details of observations and models can be found in recent reviews
\citep{WarmuthA.EIT-wave-review.2007LNP...725..107W, VrsnakCliver.corona-shock-review.2008SoPh..253..215V,
Wills-Davey.EIT-wave-review.2009SSRv..149..325W, GallagherP.LongD.EIT.wave.review2010SSRev}.

	%circular fronts of EUV enhancement followed by dimming 

	%--- STEREO, AIA new obs. ---
Major drawbacks that hindered our understanding of such propagating EUV disturbances	%EUV waves 
were EIT's single view point and low cadence ($\geq$12 minutes),	%, spatial resolution, and wavelength coverage,
which were partially alleviated by \stereo EUVI
\citep{LongD.EUVI-wave.2008ApJ...680L..81L, Veronig.STEREO-EUVI-wave.2008ApJ...681L.113V, 
MaSL.EUVI-wave.2009ApJ...707..503M,
Kienreich.2009ApJ...703L.118K, Patsourakos.EUVI-fast-mode-wave.2009ApJ...700L.182P}. 
	%Patsourakos.EUVI-wave.2009SoPh..259...49P}. 
	%--- niche ---
	%The Solar Dynamics Observatory (\sdoA) launched on 2010 February 11, joined by \stereoA,
	%offers a greater chance to put the decade long debates on the nature of EUV waves to an end. 
The Atmospheric Imaging Assembly (AIA) on the Solar Dynamics Observatory (\sdoA)
has	%observes the EUV corona at 
10 EUV and UV wavelengths, 	%almost simultaneously 
covering a wide range of temperatures,
at high cadence (up to 10--20~s, $\sim$$10\times$ faster than \stereoA's 75--150~s) and resolution ($1\farcs4$, with $0\farcs6$ pixels).
These capabilities allow us to study the kinematics and thermal structure of EUV disturbances	%waves
in unprecedented detail, as reported in this Letter.

%=======================================================================================
\section{Observations and Data Analysis}
\label{sect_data}

\begin{figure*}[thbp]      %---------------------------  
						 % use [t] [h] to move all figs to end of paper, without moving them in tex file, 
 \begin{center}
 	%\epsscale{0.9}	%{1.0}
	% \plotone{f2a.eps} \plotone{f2b.eps} \plotone{f2c.eps}		% \plotone cause x shift a bit between STEREO and AIA figures, but \includegraphics ok.
 \includegraphics[width=7.2in]{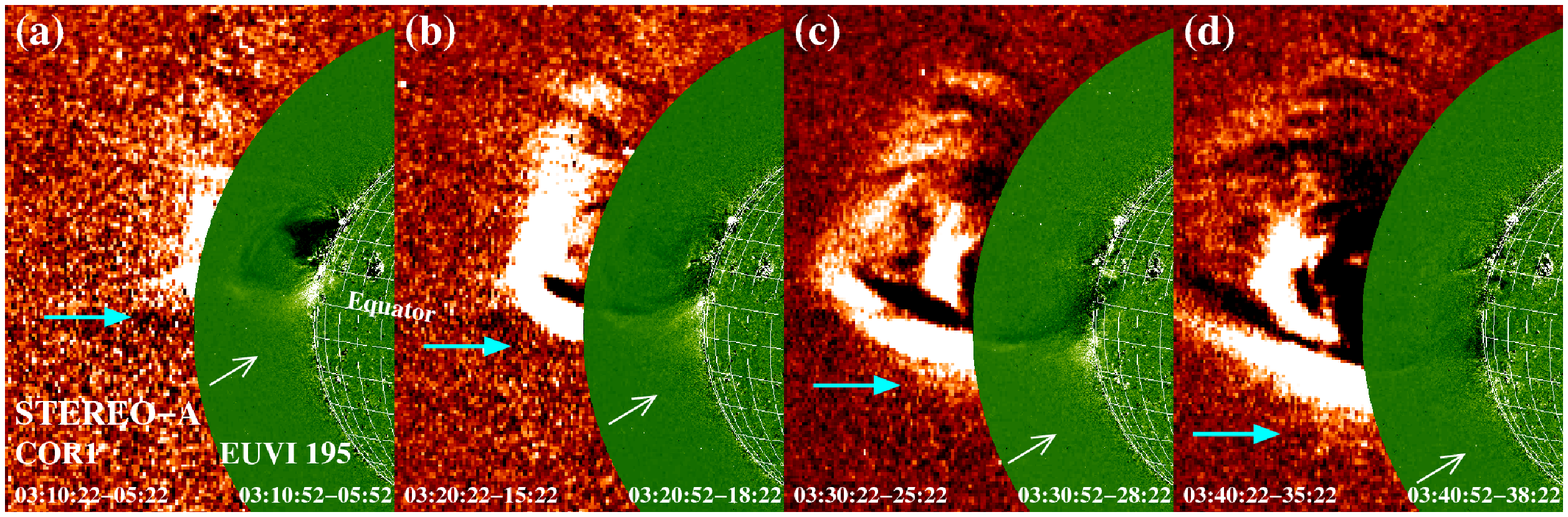}	%[width=7in]
 \includegraphics[width=7.2in]{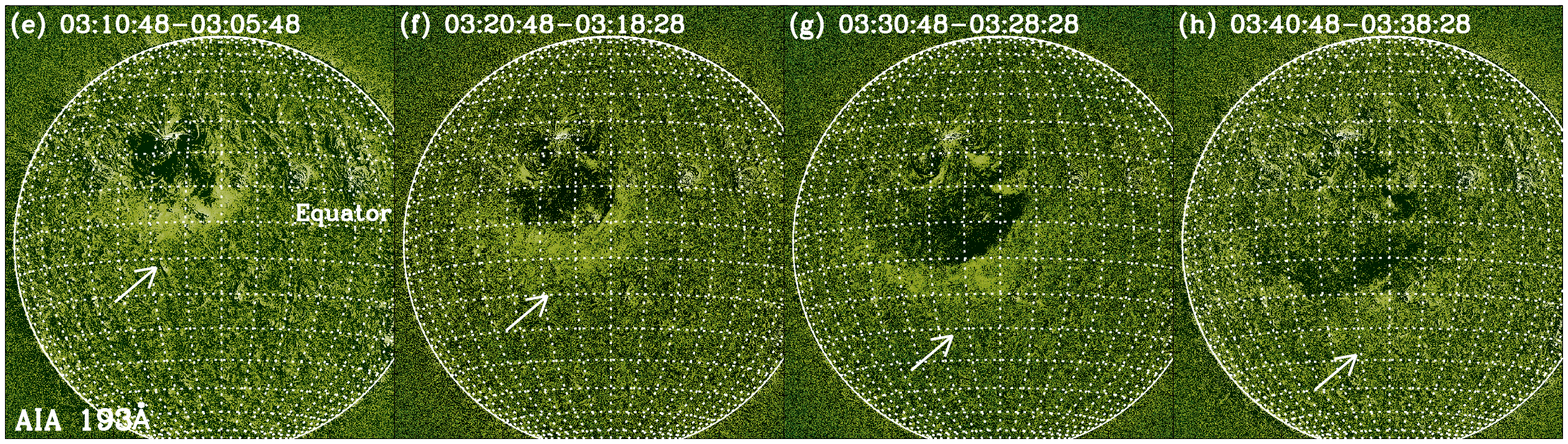}	%[width=7in]
 \caption[]{Running difference images of composite \stereoA-A COR1 and EUVI 195~\AA\ (top, see online Movie~1)	% made by the {\it Festival} software (top)
 and AIA 193~\AA\ (bottom), where the ecliptic and solar norths are up, respectively. Heliographic grids are spaced by $10\degree$.
 Arrows point to the leading front of the diffuse pulse seen in whitelight and EUV.
 The cyan arrows are also approximately the projection of AIA's line-of-sight (LOS; in the ecliptic plane) seen by \stereoA-A.
 } \label{stereo.eps}
 \end{center}
 \end{figure*}

The event of interest (Solar Object Locator: SOL2010-04-08T02:30:00L177C061) 	% SOL locator from: http://www.lmsal.com/hek/, see this event (flare) summary
occurred in NOAA active region (AR) 11060 (N25E16)	% (N31E17) 
from 02:30 to 04:30~UT on 2010 April 8.
It involved a \goes B3.8 two-ribbon flare, CME, and global EUV disturbance, as shown in Figure~\ref{stereo.eps}.
No Moreton wave was found in available \Ha data (from the Huairou Solar Observing Station, China).
%

	%---------------------------------------------------------------------------------
	%\subsection{Data Reduction}
	%\label{subsect_data}

	%instrument issues, cadence - could be shorter to save space
	%--- pointing, coalignment ---
	%see /aia/pointing/run02_detrend/193/ref-img1_eit

\begin{figure*}[thbp]     %[t] ---------------------------  
						 % use [h] to move all figs to end of paper, without moving them in tex file, 
 \epsscale{1.05}	%{1}{1.1}
 \plotone{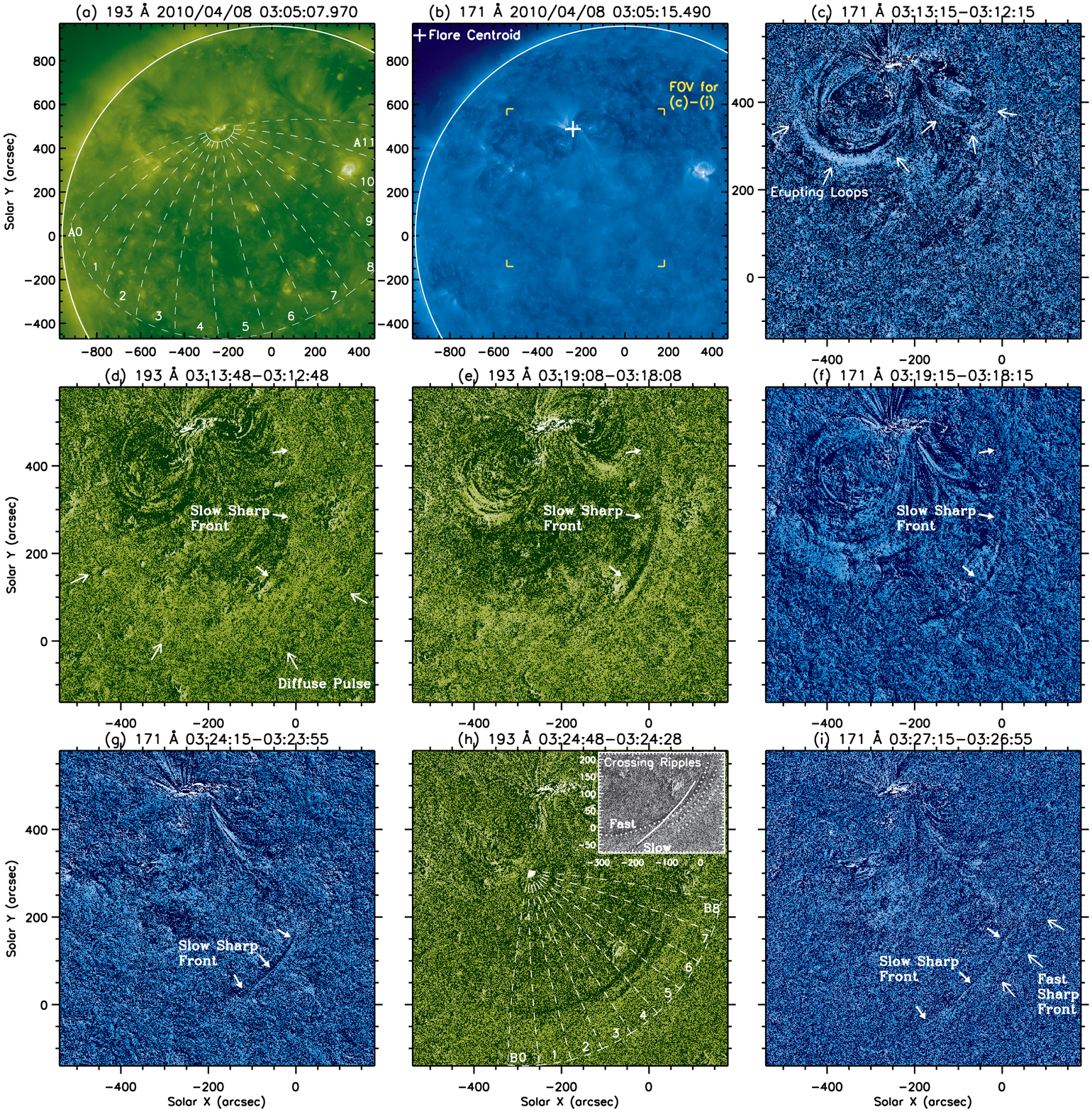}
 \caption[]{	%Development of the event at AIA images at six wavelengths: 
 Sequential AIA images at 193 and 171 \AA: (a)--(b) original (see online Movies~2a--b) 
 and (c)--(i) running difference (Movies~2c--d), with two field of views (FOVs).
 Overlaid are two sets of spherical sectors, A0--A11 in (a) and B0--B8 in (h),
 used to obtain space-time plots.	
	% Arrows in (c)--(f) are at fixed locations.
 The insert in (h) highlights multiple ripples.
 } \label{mosaic.eps}
 \end{figure*}

AIA images (20~s cadence) were first %carefully 
coaligned across wavelengths and differentially rotated to a reference time (03:33:47~UT).
Figure~\ref{mosaic.eps} shows selected images.
To minimize human subjectivity, we employed a semi-automatic approach
\citep[cf.,][]{Wills-Davey.auto-detect-EIT-wave.2006ApJ...645..757W}, following 	%the idea of 
\citet{Podladchikova.NEMO.2005SoPh..228..265P}.
We identified the eruption center (flare kernel centroid	%with the centroid of the sheared flare kernel (ribbons) in a selected 193~\AA\ image at 02:43:26~UT
at $x=-238\arcsec$, $y=487\arcsec$)		%, Figure~\ref{mosaic.eps}(b))	%heliocentric coordinates: 
	%N$30.6\degree$, E$16.7\degree$ or .
	%data at /aia/wave/run03_postSPD/193_run0/wave_center
as the new ``north pole", and drew  $15 \degree$ wide heliographic ``longitude"
sectors ({\bf A0--A11}, Figures~\ref{mosaic.eps}(a)--(b)). For each sector,
we obtained the image profile as a function of distance	% (in arcsecond or 726~km) 
measured from the eruption center along ``longitudinal" great circles 
(thus correcting for the atmosphere's sphericity)	%projection effects), 	%on the photosphere,
by averaging pixels in the ``latitudinal" direction. 
%
%In doing so, we effectively projected all coronal features of unknown heights onto the photosphere along the line of sight (LOS). 
%Uncertainties resulting from this assumption are negligibly small, is $<$1\%. 
%
Composing such 1D profiles over time gives a 2D space-time plot,	% (e.g., Figure~\ref{sliceA.eps}),
from which 	%to which additional analysis, including base and running differences and ratios 
base and running ratios and differences (giving different color scaling
suited for different features) can be readily obtained, as shown in Figures~\ref{sliceA.eps} and \ref{sliceB.eps}.
%

	%--- 2) EIT wave: cross-wavelength simple overview ---
The global disturbance is visible in all seven AIA EUV channels
and for the first time observed at 131, 94, 335, and 211~\AA.
It is similar in channels $>$1~MK, being the strongest at 193 and 211~\AA. 
It appears different in the cooler 171~\AA\ channel	and the weakest at 304 and 131~\AA.
Here we focus on the representative 193 and 171~\AA\ channels (hereafter 193 and 171).
Incidentally, we noticed very fast (500--$1200 \kmps$), narrow-angle features launched southward from the erupting AR,
repeatedly at $\sim$100~s intervals throughout a 1.5~hrs duration,		% as tentative evidence of {\bf fast mode} MHD waves.
even after the global disturbance had left.
Their nature is under investigation and will be presented in the future.

	%--- (fig. 1) event time-line, EIT wave: cross-wavelength simple overview, then focus on 193 and 171 ---

%---------------------------------------------------------------------------------
%=======================================================================================
\section{Erupting Loops and CME}		%: Possible Driver of Disturbance}
\label{sect_loop}	%{subsect_loop}

 \begin{figure*}[thbp]      %---------------------------  
 \epsscale{1.15}	%{1}{1.15}
 \plotone{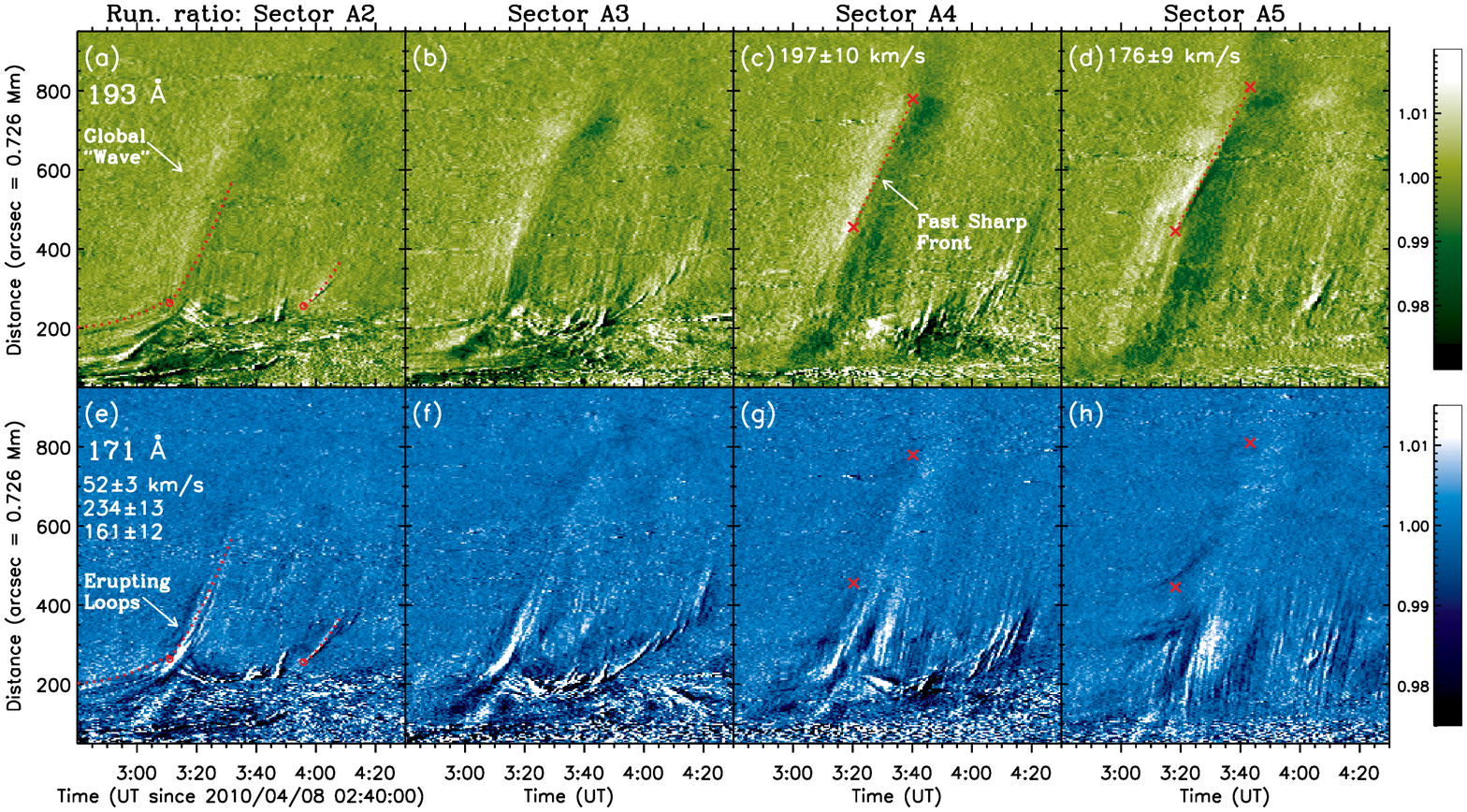}
 \caption[]{Running ratio space-time plots obtained from 
 A~sectors shown in Figure~\ref{mosaic.eps}(a) at 193 (top) and 171 (bottom), smoothed with a 60~s boxcar.
 In (c) and (d), cross signs connected by dotted lines mark linear fits to the peak of the
 diffuse pulse at 193,
 with the fitted velocities printed on the top.
 Their end points are repeated in the corresponding 171 panels at the bottom.
 Other dotted lines are	parabolic (linear in Figure~\ref{sliceB.eps}) fits to various features,
 with the greater of the initial and final velocities listed on the left in the order of the start time of the fit 
 (marked by an open circle). Fits in (e) are repeated in (a).
 Movie~3 shows this figure with and without labels and fits.
 } \label{sliceA.eps}
 \end{figure*}
%
	%--- 1) expanding loops ---
	%In the erupting AR 11060,
There is initially a series of coronal loops, best seen at 171,
fanning out along an arcade in the northwest-southeast direction. 
	%The sheared flare ribbons and post-flare loops suggest that the pre-eruption loops are also likely to be highly sheared. 
At the flare onset of 02:32~UT, 	%simultaneously with the flare onset, 
	%loops at both ends of the arcade, particularly in the southeast, 
these loops, as shown in Figure~\ref{mosaic.eps}(c), start to expand southward to the quiet Sun,
likely avoiding the probably stronger magnetic field of large scale loops in the north
(see Movies~2a--b).
	%see time slice data 171, sector A0-2, /aia/wave/run03_postSPD/171_run0/track-wave/run1_rdiff-trk
This first occurs in the inner core of the AR and successively proceeds to surrounding loops.
%
	%--- 2) 171 acceleration - CME driven; ---
The expansion experiences a gradual and then rapid acceleration. 
Parabolic fits to its 	%loop position in the 
space-time position (Figure~\ref{sliceA.eps}(e)) before and after 03:11~UT yield 
accelerations of $26\pm 1$ and $94 \pm 5 \mpss$		%sqrt(fit error^2 + 5%	system error^2)	%$26\pm 1$ and $2280 \pm 150 \mpss$ 
and final velocities of $52 \pm 3$ and $234 \pm 13 \kmps$, close to the
286~$\km \ps$ velocity of the halo CME measured by \soho LASCO.
	%data from /aia/wave/run03_postSPD/171_run0/track-wave/run1_rdiff-trk/sect02/023701 and /031055
\stereo Ahead, $67\degree$ east of the Earth, observes this event at near quadrature	%nearly $90\degree$
and indicates that the lateral expansion of the CME is also primarily toward the south
(Figure~\ref{stereo.eps}, {\it top}; Movie~1). The similar velocities and common direction
suggest that the erupting loops are part of the CME body.
 %$67\degree$ Ahead ($72\degree$ Behind) the Earth at 03:10~UT (03:15).
 %	%data: see Nariaki Festival STEREO movies

%=======================================================================================
\section{Global EUV Disturbance}	%: Results and Discoveries}
\label{sect_result}

%---------------------------------------------------------------------------------
%\subsection{Global EUV ``Wave"}
%\label{subsect_eit.wave}

%---------------------------------------------------------------------------------
\subsection{Overview}	%{Global EUV Disturbance Overview
\label{subsect_global}
	%\subsubsection{Coherent Front at Constant Speed}	%\label{subsubsect_constv}

	%--- two components: diffuse vs. sharp (slow and fast): mention here first and then refer to later ---
The global EUV disturbance, seen by AIA on the solar disk, exhibits two components: one broad, {\bf diffuse pulse}
superimposed on which are multiple narrow, {\bf sharp fronts} (see Movies~2c--d). 
The diffuse pulse appears as an 	%doughnut
annular-shaped emission enhancement ahead of the erupting loops.
It expands in all directions except to the north, being the strongest in the southwest 
(Figure~\ref{stereo.eps}, {\it bottom}; Figure~\ref{mosaic.eps}(d)).
The sharp fronts	%, best seen in running difference images,
also have two components: a {\bf slow front} that appears earlier
	%(first seen at 03:14~UT, a circular arc of the global EUV wave front is first seen at
	%$400\arcsec$ (290~Mm) to the southwest of the flare kernel) 
and occupies from the west to southwest of the AR
(Figures~\ref{mosaic.eps}(d)--(g)), and a {\bf fast front} that comes later near the peak location of the 
diffuse pulse and dominates from southwest to south (Figures~\ref{mosaic.eps}(h)--(i)).

	%--- (fig. 2) wave formation and propagation ---
	%--- 1) vel in different directions ---
In running ratio	% (similar to time derivative)
space-time plots at 193 (Figure~\ref{sliceA.eps}, {\it top}),		% obtained from A sectors shown in Figure~\ref{mosaic.eps}(a).
the diffuse pulse appears as a diagonal stripe of emission increase (bright) followed by decrease (dark),
the boundary between which corresponds to the pulse peak.	% in a large angular range.
In Sector~A4, for example, the peak coincides with the fast sharp front 	%in this case, 
and propagates 	%at a constant velocity of $197 \pm 10 \kmps$ for more than 20~minutes	% (03:20--03:40~UT) %from $400 \arcsec$ 
up to $800 \arcsec$ (580~Mm) from the flare kernel.
At 171 (Figure~\ref{sliceA.eps}, {\it bottom}), the pulse is less pronounced and, in contrast, appears
as emission {\it reduction} \citep[dark followed by bright;][]
{Wills-DaveyThompson.TRACE-EUV-wave.1999SoPh..190..467W, DaiY.EUVI-wave.non-wave.2010ApJ...708..913D}
rather than commonly observed {\it enhancement} \citep[e.g.,][]{LongD.EUVI-wave.2008ApJ...680L..81L}. 
Together, these observations suggest that the cooler 171 material is heated to 193 temperatures.

	%--- compare direction and position between ST-A and AIA for EUV wave. ---
The side view from \stereoA-A indicates that the diffuse EUV enhancement is confined in the low corona
and its leading front matches the lateral extent of the whitelight CME	% seen by COR1
\citep{	%Patsourakos.EUVI-wave.2009SoPh..259...49P, 
Patsourakos.EUVI-fast-mode-wave.2009ApJ...700L.182P, Kienreich.2009ApJ...703L.118K}.
It is followed by an arc-shaped sharp increase and then decrease of emission that
extends into the COR1 FOV (high corona).	%, similar to what has been seen.	% We speculate that 
Its projection onto the disk likely corresponds to the diffuse pulse's peak 
and/or one of the sharp fronts seen by AIA, according to its latitudinal position,
sharpness, and large vertical extent (thus large LOS integration).
	%and position behind the diffuse front.

%---------------------------------------------------------------------------------
%\newpage
\subsection{Diffuse Pulse Shape Evolution}
\label{subsect_pulse}

 \begin{figure*}[thbp]      %---------------------------  
 \epsscale{1.05}	%{1}{1.15}
 \plotone{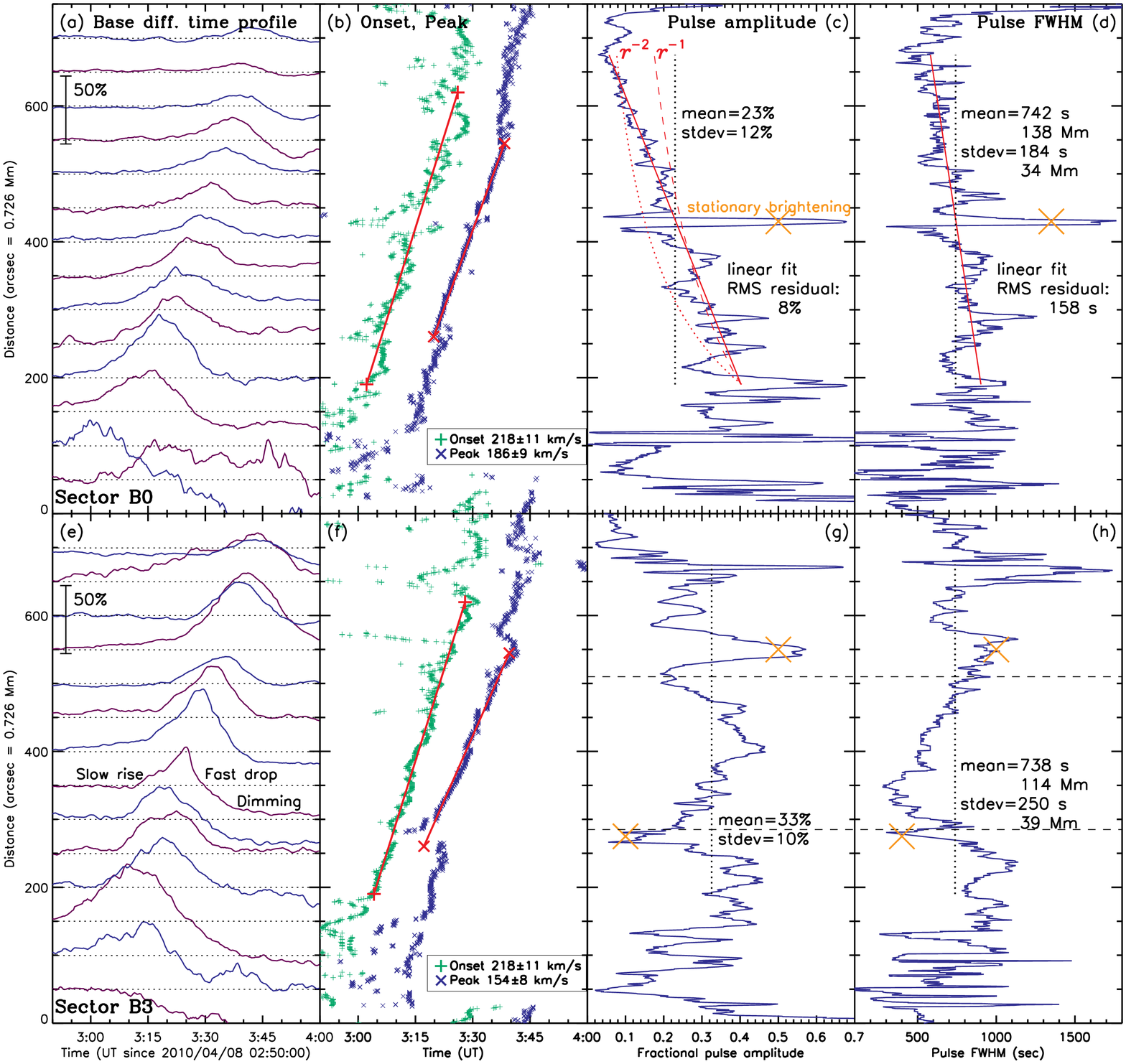}
 \caption[]{	%Closeup of 193 disturbance propagation for Sectors B0 (top) and B3 (bottom).
 {\it Top}: (a) Horizontal cuts of the normalized 193 base difference space-time plot of Sector B0 shown in Figure~\ref{mosaic.eps}(h).
 Each curve represents the percentage emission variation at a fixed location marked by the
 dotted line (also representing the pre-event level). 
	% The vertical bars indicate the scale of a 50\% increase. 
  (b)--(d) Distance vs.~onset and peak times, amplitude, and FWHM of the	%intensity enhancement
 pulse shown in (a). 
 The red solid lines are linear fits, and the red dashed and dotted lines in (c) are $r^{-1}$ and $r^{-2}$ fits
 starting with the same value at $s=190\arcsec$.
 Vertical dotted lines mark the mean values in the distance range of interest.
 The orange ``$\times$"s in (c)--(d) indicate stationary brightenings of small scale loops.
	% (e)--(h) Same as (a)--(d) 
 {\it Bottom}: same as {\it Top} but for Sector B3.
 } \label{slice-prof.eps}
 \end{figure*}
To best track the strong disturbance in the southwest, we used a finer set of $10\degree$ wide sectors 
({\bf B0--B8}, Figure~\ref{mosaic.eps}(h)) centered at $x=-275\arcsec$, $y=306\arcsec$, 
which are approximately perpendicular to the local disturbance front (to which the A~sectors are not).
	%of the wave front arc at 03:24:26~UT and 193~\AA\ (see Figure~\ref{mosaic.eps}(h)).
	%data at /aia/wave/run04_ripple/193/get-cut

For selected Sectors~B0 and B3, we composed base difference space-time plots, 
normalized by their average pre-event brightness. We then obtained time profiles 
from horizontal slices at every distance of the space-time plot.
As shown in Figures~\ref{slice-prof.eps}(a) and (e) for 193, the profiles represent net emission 
changes and clearly show a single hump successively delayed with distance,
corresponding to the propagating diffuse pulse. 
We define the pulse onset at the 20\% level between the peak and the pre-event minimum,
from which the pulse amplitude and FWHM are calculated.
 %their amplitudes are directly comparable at different positions 
 %\citep[cf.,][]{Veronig.dome-wave.2010ApJ...716L..57V}. % for short
 %Delayed occurrences of the pulse at increasing distances manifest its propagation.
	%Note that perturbation profiles obtained from running ratio images \citet{Veronig.dome-wave.2010ApJ...716L..57V}
	%cannot be used for quantitative analysis, since they have different normalizations in both space and time.
	%As we have seen, these simple pulses are fairly stable and propagate at constant velocities.
Note that these temporal profiles at fixed locations are equivalent to spatial profiles at given times
(i.e., vertical slices of space-time plots), because the pulse is fairly stable and propagates 
at constant velocities (see Section~\ref{subsect_kinem}).

	%--- B0: simple, no multiple ripple, no crossing ripples; ---
In B0, where the slow sharp front is absent,	% and no front overtaking occurs,
the diffuse pulse clearly develops beyond $s_0=190\arcsec$ ($370\arcsec$ south of the flare kernel),
within which it is not well-defined.
	%after the initial irregularities. 	%at short distances, 
 %It has a gradual rise and drop, except for a mildly sharp rise near $350\arcsec$.		% during the middle stage
	% which is far from steep enough to form a shock. 
The pulse amplitude can be fitted with a linear function, which decreases with distance 
by a fraction of ${6 \over 7}$ from a 41\% enhancement at $s_0$ to 6\% at $675\arcsec$
(Figure~\ref{slice-prof.eps}(c)).	%RMS residual of 8\% for the fit. 
	%(The spike near $430\arcsec$ is due to brightening of small-scale loops.)	% coronal loops on the path.)
(For comparison, the corresponding maximum emission reduction at 171 is only $\sim$10\%.)
It decreases faster than the corresponding $r^{-1}$ fit but slower than $r^{-2}$, 
implying the pulse being neither a surface nor a spherical wave.
The pulse FWHM decreases by only ${1 \over 3}$ from 902 to 581~s		%36\%
(Figure~\ref{slice-prof.eps}(d)), 	%with a 158~s residual for the linear fit.
rather than increases as expected from dispersion of a wave. % for short.
	%data, /aia/wave/run04_ripple/193/track-wave/run0_auto-peak/bdif_sect0/vfit2_zmax/fit_time_slice_peak.dat, /sect0/vfit3_fwhm
	%and normalization from ../base_ave_norm.dat; print, [115.018, 16.3232, 21.3614]/283.42354= 0.405817, 0.0575930, 0.0753692 (max, min, RMS residual)

	%--- B3: complex, interaction ripples, sharp drop - superposition or interference.
In B3, 	%only $20 \degree$ away from B0,
where the fast and slow sharp fronts overlap (see Movie~4b), the pulse shape is quite different.		% and has larger variations, lacking a general trend. 
Despite complications caused by stationary brightenings or dimmings of small-scale loops,
	%a weak anti-correlation is found in the distance range marked by the dashed lines.	% $s=285 \arcsec$ and $510 \arcsec$.
an increase followed by a decrease of the pulse amplitude is roughly anti-correlated with
variations of the FWHM in the distance range marked by the dashed lines 
(Figures~\ref{slice-prof.eps}(g)--(h)). 
This indicates steepening and then broadening of the pulse. 
The narrowest pulse (near $s=350\arcsec$) shows a sharp
drop and deep, long-lasting dimming on its trailing edge, 
which are different from leading edge steepening in shock formations.

	%Such steepening seems to be related to the interaction of the
	%fast packet overtaking the slow front mentioned above, which will be further discussed later. 

%---------------------------------------------------------------------------------
\subsection{Kinematics of Disturbance Propagation}
\label{subsect_kinem}

\begin{figure*}[thbp]      %---------------------------  
 \epsscale{1.15}	%{1}{1.15}
 \plotone{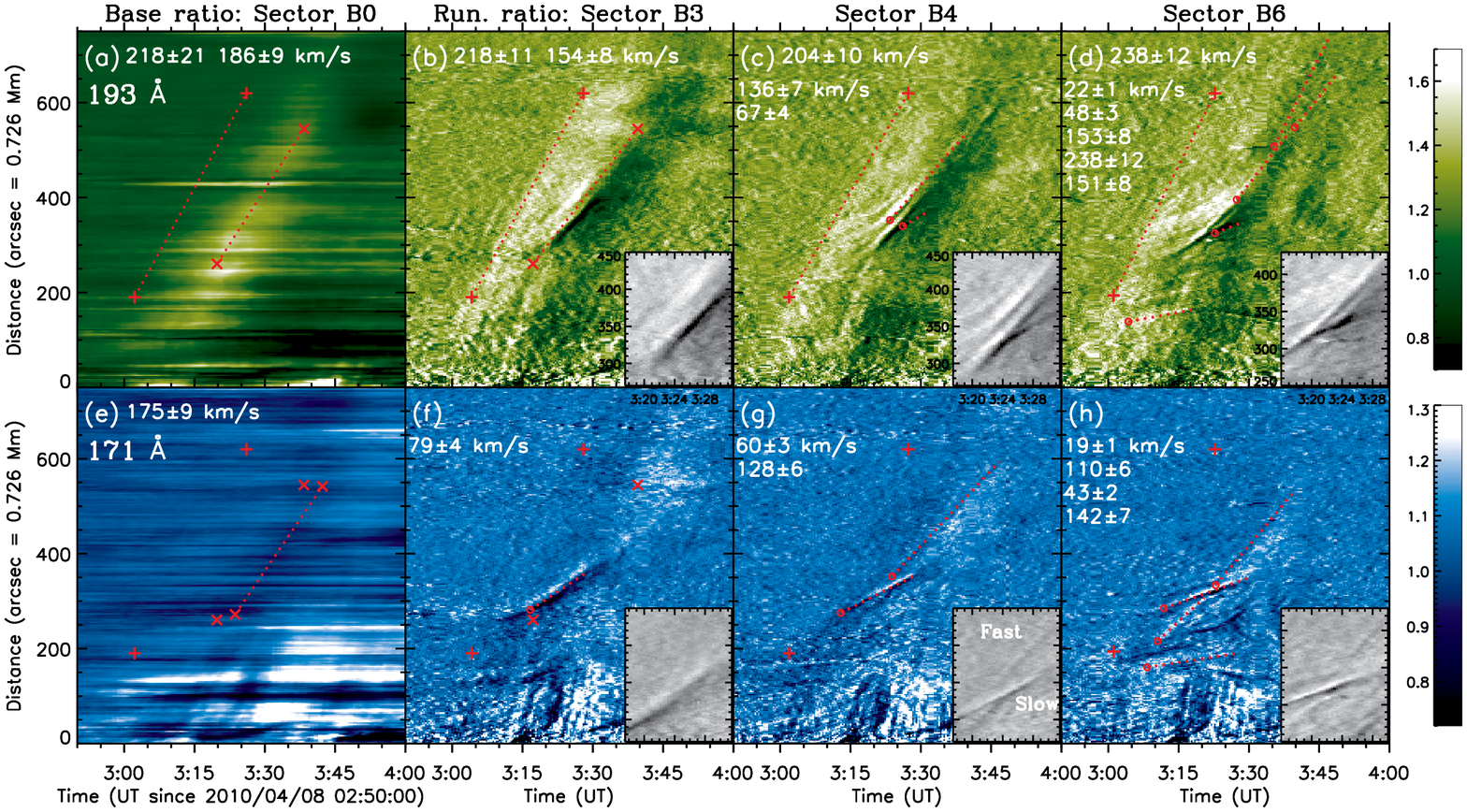}
 \caption[]{
 Same as Figure~\ref{sliceA.eps} but for base (column~1) and running (columns~2--4) ratios 
 from B~sectors shown in Figure~\ref{mosaic.eps}(h).
 Plus (cross) signs	% $+$'s ($\times$'s)
 connected by dotted lines 	%in the first two columns 
 are linear fits to the onset (peak)	%(peak for 193, trough for 171) 
 of the diffuse pulse.	%, with the fitted velocities marked on the top. 
 Their end points for 193 are repeated for 171 at the bottom. 
 The color bars on the right are for column~1 only, while columns~2--4 share the color scales with Figure~\ref{sliceA.eps}.
 The gray scale inserts provide enlarged views showing parallel and crossing ripples.
 (See online Movies~4a--c for this figure and running difference images from which the space-time plots are
 obtained.)
 } \label{sliceB.eps}
 \end{figure*}

We derive kinematics of the disturbance propagation from
base and running ratio space-time plots for B sectors as shown in Figure~\ref{sliceB.eps}.

	%---------------------------------------------------------------------------------
	%\subsection{Detailed Wave Properties}	%{Closeup: Overtaking Ripples, Coherent Pulse}
	%\label{subsect_closeup}
	%\subsubsection{Fine Structure: Multiple Ripples}	%\label{subsubsect_ripples}

%
%----------------------------------------------------------
\subsubsection{Diffuse Pulse}
\label{subsubsect_diffuse}

	%--- 1) 193: overview ---
At 193 (Figures~\ref{slice-prof.eps}(b) \& (f) and \ref{sliceB.eps}(a)--(d)), both the onset and peak of the diffuse pulse travel at constant velocities in their mid stages, 
with the former being slightly faster (e.g., $218 \pm 21$ vs.~$186 \pm 9 \kmps$ for Sector B0).
This, despite the lack of a general trend for the overall pulse FWHM noted above,
indicates that the rising portion of the pulse broadens at increasing distances in all directions,
consistent with that found by \citet{Veronig.dome-wave.2010ApJ...716L..57V}.
In addition, the pulse onset velocity weakly depends on directions, with a narrow range of $204$--$238 \, \kmps$.

	%--- 1) 171 wave kinematics ---
At 171 (Figures~\ref{sliceB.eps}(e)--(h)), with the signal being considerably weaker,
the maximum emission reduction of the diffuse pulse propagates at similar velocities as its 193 counterpart of emission enhancement,
e.g., $175 \pm 9$ vs.~$186 \pm 9 \kmps$ for Sector~B0.
There is a delay by $\sim$230~s in time or $\sim$42~Mm	%83~Mm => wrong
in space, which we speculate could result from different wavelength response to the convolution
of temperature and density effects, both being likely present here.
	%data: /aia/wave/run04_ripple/171/track-wave/run0_auto-peak/sect0/vfit0_peak/fit_time_slice_peak.dat
	%IDL> print, 3722.11-3492.24 =>     229.870 sec

%----------------------------------------------------------
\subsubsection{Sharp Fronts}
\label{subsubsect_sharp}

	%--- hand-fit individual ripple/ridges -----
At 193, the sharp fronts are recognized as narrow ridges 
in running ratio space-time plots (Figures~\ref{sliceB.eps}(b)--(d)),
and some of them experience acceleration (e.g., Figure~\ref{sliceB.eps}(d)).
Their final velocities in different directions vary by a factor of two ($136$--$238 \kmps$), 
in contrast to the narrow range of the diffuse pulse's onset velocity. %leading edge.

	%--- 2) 171 fine front: bright (not dark) ---
At 171 (Figure~\ref{sliceB.eps}, {\it bottom}), most of the sharp fronts are emission enhancements, opposite to the diffuse pulse reduction,
but the same as their 193 counterparts at similar velocities 
(e.g., $128 \pm 6$ vs.~$136 \pm 7 \kmps$ for B4).
This suggests a density increase at both channels' temperatures.
The slow sharp front here is particularly pronounced and dominant over the fast front
(see Figures~\ref{sliceB.eps}(f)--(h), inserts).
 %(Yet it is somewhat different at 193 in spatial extent, brightness, and kinematics.)
Its velocity is only $\sim$50\% that of the diffuse pulse's peak (e.g., $79 \pm 4$ vs.~$154 \pm 8 \kmps$
for B3) and decreases toward the west,	% ($43 \pm 2 \kmps$ for B6).
giving rise to its southward elongated oval shape (Figure~\ref{mosaic.eps}(f)).

	%--- slow 20-50 km/s: expanding/imploding loops ---
Incidentally, there are numerous slow ($\sim$$20 \kmps$), outward moving features
(B6, Figure~\ref{sliceB.eps}), which are likely the flanks of the expanding loops 	%undergoing expansion and then implosion
mentioned earlier. The onsets 	%of these features as well as those accelerating ripples 
of these features and the slow sharp front occur within minutes of the arrival of the diffuse pulse's
leading edge at increasing distances, suggesting the former being triggered by the latter.  

	%two separate questions: 1) 171 feature is enhancement or reduction, 2) cospatial with what at 193?

%---------------------------------------------------------------------------------
\subsection{Component Interaction: Crossing and Parallel Ripples}
\label{subsect_ripple}

Near 03:24~UT, the fast sharp front, traveling two times faster, overtakes the
slow sharp front initially ahead of it at a shallow angle (see Figure~\ref{mosaic.eps}(h), Movies~4b--c).
They appear as crossing sharp ridges in space-time plots (Figure~\ref{sliceB.eps}, inserts)
and both propagate independently afterwards.

	%--- parallel ripples ---
	%Around this time, a
Two additional weaker, sharp fronts (``{\bf ripples}") are generated ahead of
the original fast front (Figures~\ref{mosaic.eps}(h) and \ref{sliceB.eps}(c)).
Early in their lifetime 	%(03:18--03:22~UT), 
the additional fronts accelerate, asymptotically approaching the constant velocity 
of the original front, while their spatial separations decrease with time.
Afterwards, they propagate at the same constant velocity of
$\sim$$130 \kmps$ for up to $\sim$10~minutes, maintaining a constant separation of 
$\sim$160~s in time or $\sim$20~Mm in space. 
This the first time such fine ripples are observed and they are different from those more widely separated
(by fractions of $\Rsun$) \ion{He}{1} ``wave" pulses \citep[][their Fig.~2]{Gilbert.HeI.wave.2004ApJ...610..572G}.
	%are more widely separated in space (fraction of $\Rsun$) 
	%and time ($\sim$10 minutes).

	% click to measure from plotman
	%IDL> print, anytim('03:27:56.464') - anytim('03:25:16.240') =   160.22400 sec
	%IDL> print, 160.22400*130./1e3	=      20.8291 Mm

%---------------------------------------------------------------------------------
%\subsection{Dimming and Brightening}
%\label{subsect_dim}

%---------------------------------------------------------------------------------
%\section{Repeated Fast ($\sim$1000~$\km \ps$) Features}
%\label{sect_fast}

%=======================================================================================
\section{Discussion}	%Summary and Discussion Conclusion
\label{sect_discuss}	%{sect_conclude}

	%We have presented the 
The first \sdo AIA observations of a global EUV disturbance presented here
have revealed a picture of multiple temperatures and spatio-temporal scales
in unprecedented detail. 
	%AIA's high cadences and sensitivities led to the discoveries of 
	%the two components of diffuse and sharp fronts and crossing ripples.	
	%, and repeated $\sim$$1000 \kmps$ fast features.
	% were made possible for the first time
	%Many of these fine structures could be missed by previous instruments, even at EUVI's highest 75~s cadence.

%---------------------------------------------------------------------------------
\subsection{Separation of Sharp Fronts from Diffuse Pulse}	%{Two Components of Diffuse and Sharp Fronts}
\label{subsect_2component}

	%--- emphasize difference from "S" wave, Harra & Alphonse ---
AIA's high cadences and sensitivities allowed us to discover
and separate a new feature, the sharp fronts, from the concurrent, commonly seen diffuse pulse.	% and crossing ripples. 
With overtaking fast and slow components and no Moreton wave association, the sharp fronts 
	%which are temporally unresolved with EIT or EUVI. 
seem to differ from ``S-waves", a minority \citep[$\sim$7\% of 173 events;][]{Biesecker.EIT-wave-association.2002ApJ...569.1009B} 
of ``EIT waves", that are sharp, fast, and often cospatial with Moreton waves 
\citep{ThompsonB.EIT-Moreton-wave.2000SoPh..193..161T}. 

	%--- cannot say this: division between bright and dark is not the trailing edge of diffuse wave -----
	%--- but close to its peak, like in our case
	% => best avoid debate w/ Meredith, don't mention Harra's paper here, but later for 500 km/s ok -----
	%They also differ from the ``bright wave" \citep{HarraSterling.fast.leading.edge.EIT-wave.2003ApJ...587..429H}
	%identified at the trailing edge of a diffuse pulse in running difference images.

%---------------------------------------------------------------------------------
\subsection{Hybrid Wave and Non-wave Hypothesis}
\label{subsect_hybrid}

%As positive and negative evidence has been found in this event for both the
%wave and non-wave models of global EUV disturbances,
	%A hybrid model combining both mechanisms may provide the best, although not entirely satisfactory, 
	%explanation. 
We propose a hybrid hypothesis combining both wave and non-wave aspects
to best, although not entirely satisfactorily, explain the observations.
Namely, the sharp bright fronts are caused by CME compression traveling behind the
weaker, more uniform diffuse front which is an MHD wave generated by the CME.
This is conceptually similar to the wave/non-wave bimodality 	%of wave and eruptive modes
suggested by \citet{Zhukov.EIT-wave2004A&A...427..705Z} and simulated by
\cite{CohenO.EITwave.non-wave.both.2009ApJ...705..587C}.
Our specific interpretations are as follows.

\begin{enumerate}	%====================

%--- sharp fronts: ---------

 \item 	%--- location ---
The bright sharp fronts are located to the southwest of the erupting AR, consistent with the main direction of the CME expansion,
which likely produces the strongest compression. 

 \item 	%--- acceleration ---
Some sharp fronts exhibit accelerations and their final velocities (e.g., $238 \pm 12 \kmps$, Figure~\ref{sliceB.eps}(d))
are close to those of the erupting loops and CME (Section~\ref{sect_loop}). Such accelerations
\citep{Zhukov.EIT-wave-acc-decel.2009SoPh..259...73Z}, not expected for freely propagating MHD waves, 
likely reflect the acceleration of the CME in the low corona \citep{ZhangJie.CME-flare-C1.2001ApJ...559..452Z}.

 \item 	%--- compression ---
The emission profile steepens on the trailing edge of the sharp front (Figure~\ref{slice-prof.eps}(e)), suggestive of compression
by the CME from behind. It is followed by deep dimming, another indicator of the expanding CME body.
Such compression is also consistent with the density enhancement suggested by brightening of the sharp fronts at both 193 and 171.

%--- diffuse pulse ---------

 \item 	%--- weak, location, velocity ---
The leading edge of the diffuse pulse	%, being weaker in brightness,
travels at constant velocities with very little direction-dependence. 
This is expected for a fast mode MHD wave. 		% propagating at the characteristic speed of a uniform medium. 
Its velocity of 204--$238 \kmps$ \citep[cf., $500 \kmps$,][]{HarraSterling.fast.leading.edge.EIT-wave.2003ApJ...587..429H} 
is roughly within the estimated range of fast mode speeds on the quiet Sun
\citep[$215$--$1500 \kmps$,][]{Wills-Davey.auto-detect-EIT-wave.2006ApJ...645..757W}.
In the absence of a shock, this speed should be greater than that of the wave generator (i.e., the CME
preceded by the sharp fronts), as observed here. 

 \item 	%--- pulse shape ---
The linear drop of the pulse amplitude and the narrowing of the pulse width with distance
	%The linear decrease with distance of the amplitude and FWHM of the overall pulse
\citep[Section~\ref{subsect_pulse}; cf.][]{Veronig.dome-wave.2010ApJ...716L..57V}
 are inconsistent with any freely propagating MHD waves, 
but might be explained by the superposition of the CME-driven wave and non-wave components.

\end{enumerate}		%====================

%---------------------------------------------------------------------------------
\subsection{Open Questions}
\label{subsect_question}

The main challenge to the above hybrid interpretation	%, as well as to both wave and non-wave models, 
is the interaction of the fast and slow sharp fronts (Section~\ref{subsect_ripple}),
which shows many characteristics of real waves.		%, including crossing and parallel ``ripples".
	%Some features could be alternatively explained by the expanding CME. % \item	%--- parallel ripples ---
	%For example, the secondary, accelerating ripples forming ahead of the original fast
	%sharp front (Figure~\ref{sliceB.eps}(c)) could reflect multiple layers (or loops) of a
	%CME body undergoing successive acceleration.
 	%\item	%--- anticorrelation: amplitude/FWHM ---
	%However, some other features cannot be accommodated by the hybrid model.
  %\item	%--- ripple crossing - supports wave models
The overtaking fronts can be explained by two wave pulses with different velocities.
	%which can superimpose to increase the resultant amplitude and propagate independently after the
	%faster one overtakes the slower one. 
When a faster pulse approaches and overtakes a slower pulse, their spatial superimposition 		%of their spatial overlaps
naturally leads to the observed anti-correlation of the pulse amplitude and FWHM (Figure~\ref{slice-prof.eps}(g)--(h)).
They then propagate independently afterwards, as observed here.
However, if a sharp front represents a layer/loop of the 
erupting CME structure, a faster layer would sweep and compress a slower layer ahead of it to form a single layer
of stronger compression. This would not lead to overtaking, unless they are at different altitudes 
and LOS projection gives an impression of overtaking, 
for which we found no evidence from \stereo observations.

  % \item	%--- slow front too slow for fast mode.
Meanwhile, this also challenges pure wave models.
If the fast and slow sharp fronts are of the same MHD wave mode, they would propagate
at the same characteristic speed dictated by the medium, contradicted by the fact that
their observed velocities differ by a factor of two even in similar directions.
The velocities of the slow sharp front are too small for fast mode waves,
while slow mode waves are not favorable candidates either because they cannot propagate
perpendicular to magnetic fields.
  
 %\end{itemize}	%{enumerate}		%====================

	%--- two possibilities for entire scenario: evidence for wave/non-wave models, plus/minus ---------
	%additional vertical compression by CME and solar surface, aside from horizontal sweeping compression.

	%--- constant vel, cf. Long (Meredith review, end of 4.1: Kienreich, Ma's const. vel, no decelerate) --------------
%we cannot say for sure early deceleration, b/c of fast features...
Another problem with the fast mode interpretation is the smaller than expected ``wave" velocity which
was ascribed to undersample by low instrument cadences \citep{LongD.EUVI-wave.2008ApJ...680L..81L}. 
	%found in one event that low cadences lead to velocity underestimates.	% of the velocity when it varies rapidly. 
AIA's high cadences here have not revealed much anticipated higher velocities, 
and it remains to be seen if this applies to more energetic events.
	%, although a statistical result remains to be seen.  $\sim$200$\kmps$
In addition, the diffuse pulse's velocity remains constant 
\citep[consistent with earlier findings; e.g.,][]{White.Thompson.Nobeyama-EIT-wave.2005ApJ...620L..63W, MaSL.EUVI-wave.2009ApJ...707..503M}, 
rather than decelerates 	%as expected from 
like a shock (not necessarily true for ``EIT waves") degenerating to a fast mode wave \citep{WarmuthA.EIT-wave-review.2007LNP...725..107W}.
%==> well, don't bother, since this is going to invoke the fast features, which we do NOT want to talk about for now.
%Note that 
%	%the deceleration found by \citet{WarmuthA.multiw-EIT-wave.2004A&A...418.1101W} in several ``S-wave" events
%	%was primarily inferred from Moreton wave data, and 
%the apparent deceleration seen in Figure~\ref{sliceA.eps}(b) seems to be a confusion
%by the repeated transient fast features mentioned earlier.	% superimposed on the diffuse pulse.

The nature of global EUV disturbances thus remains elusive. However, this is just the beginning, 
and we anticipate that continuing analysis of new data from AIA, \stereoA, and other 
instruments will provide more critical information and help eventually end the decade-long debate.
	% An in-depth multi-wavelength and -instrument study of this event including detailed thermal analysis
	%and \stereo observations	%and comparison with theoretical models are underway and 
	%is under way and will be presented in the future.

%=======================================================================================

\acknowledgments
{This work was supported by AIA contract NNG04EA00C.
	%We thank 
We thank the referee for constructive comments
	%thank Meredith Wills-Davey and Leon Ofman for fruitful discussions,
and Richard Shine, Mei Zhang, and the \sdo team for their help.
} 

	%, to Leon Ofman for fruitful discussions,
	%, Junwei Zhao, Bart DePontieu, Astrid Veronig, Markus Aschwanden, and Tom Berger
	% , Paul Boerner, Trae Winter, Richard Nightingale, AIA and HMI 

%=======================================================================================
%\clearpage

% To-do at submission
% 1) To shorten long author lists: edit ms.bbl later by hand and simply do latex, not to invoke new bibtex entry.

%{\scriptsize
%\bibliography{bib/ads_all_edit,bib/LiuW-group,bib/Liu-Wei}
%}

{\scriptsize

}

%=======================================================================================

\end{document}